\documentclass[prc,amsmath,twocolumn,showpacs,superscriptaddress]{revtex4}
\usepackage{graphicx}
\usepackage{amssymb}
\usepackage{enumerate}
\usepackage{verbatim}

\begin{document}

\title{Improved description of $^{34,36,46}$Ar(p,d) transfer reactions}

\author{F.M. Nunes}
\affiliation{National Superconducting Cyclotron Laboratory, Michigan State University, East Lansing, MI 48824, USA}
\affiliation{Department of Physics and Astronomy, Michigan State University, East Lansing, MI 48824-1321}
\author{A.~Deltuva}
\affiliation{Centro de F\'{\i}sica Nuclear da Universidade de Lisboa, P-1649-003 Lisboa, Portugal}
\author{June Hong}
\affiliation{National Superconducting Cyclotron Laboratory, Michigan State University, East Lansing, MI 48824, USA}
\affiliation{Department of Physics and Astronomy, Michigan State University, East Lansing, MI 48824-1321}

\date{\today}

\begin{abstract}
An improved description of  single neutron stripping from  $^{34,36,46}$Ar beams at 33 MeV/nucleon
by a hydrogen target is presented and the dependence on the neutron-proton asymmetry
of the spectroscopic factors is further investigated.
A finite range adiabatic model is used in the analysis and compared to previous zero range
and local energy approximations. Full three-body Faddeev calculations are performed to
estimate the error in the reaction theory. In addition, errors from the optical potentials are also evaluated.
From our new spectroscopic factors extracted from transfer, it is possible to corroborate
the neutron-proton asymmetry dependence reported from knockout measurements.
\end{abstract}

\pacs{24.10.Ht; 24.10.Eq; 25.55.Hp}

\keywords{transfer, deuteron breakup, nuclear reactions, spectroscopic factors, zero-range, local energy approximation, finite-range}

\maketitle
\section{Introduction}

Since the origin of the shell model \cite{shellmodel}, concepts such as
orbital ordering, level occupancy and magic numbers have been extremely helpful to describe nuclear properties.
In the last couple of decades, with the advances of rare isotope facilities, it has been possible
to explore shell structure away from stability and perhaps expectedly, new orbital ordering, level
occupancies and magic numbers have been revealed (e.g. the doubly magic $^{24}$O \cite{jansens}).
In this quest of understanding shell evolution as one moves toward the limits of stability,
nuclear reactions have played a central role. Examples of the ongoing intense activity are: the
systematic knockout program at the National Superconducting Cyclotron Laboratory
(NSCL) \cite{mcdaniel,fallon10,reynolds10},
the sequence of transfer measurements on unstable fission fragments at Oak Ridge National Laboratory
(e.g. \cite{sn124,ge83se85,sn132dp}), the experiments by the High Resolution
Array (HiRA) collaboration at NSCL (e.g. \cite{ar-exp}) and
the transfer program at Argonne National Laboratory on light systems (e.g. \cite{li8dp,he6dp}).
It is of paramount importance that the reaction theory used in the analysis of these reactions provide
an accurate description of the process and that theoretical uncertainties be well understood.

Traditionally, the linkage between structure and reactions has often been made
through the notion of a spectroscopic factor, a quantity directly associated with the occupancy of the
single particle levels \cite{book}. By comparing the theoretical predictions for the cross section to the
data, experimental spectroscopic factors ($S_{exp}$) can be extracted. These $S_{exp}$ can then
be compared to predictions from structure models $S_{th}$. Although both,  $S_{exp}$ and $S_{th}$, are model dependent,
since the early days of the field, the methods for extracting $S_{exp}$ were tuned such that $S_{exp}$
for adding/removing a nucleon to a closed shell nucleus agreed with the independent particle picture.
This is nicely illustrated in the systematic study presented in \cite{tsang05}.
This procedure has now been challenged by the results from $(e,e'p)$ data \cite{vj1997,kramer}.
The $(e,e'p)$ data are consistent with a reduction of proton spectroscopic factors for
many closed shell nuclei by roughly $60\%$. As shown in \cite{kramer}, the apparent discrepancy
between transfer and electron knockout can be understood by using a consistent single particle radial
behavior for the removed particle as well as an improved reaction theory.
Along those lines, a re-analysis of (d,p) and (p,d) data on Ca isotopes, using Hartree-Fock densities
to constrain the single particle states, produce spectroscopic factors similar to those from $(e,e'p)$ \cite{lee06}.

With the possibility to study systems with large proton/neutron asymmetry, knockout studies revealed a strong dependence of the reduction factor $R=S_{exp}/S_{th}$ on the relative energy between valence neutrons and valence protons (e.g. \cite{gade08}). This exciting work has been systematized through plotting  the reduction factor $R$ as a function of the difference between separation
energies: $\Delta S=S_p-S_n (S_n-S_p)$ for proton (neutron) knockout. In this plot,
all the stable systems previously studied through $(e,e'p)$  correspond to $\Delta S \approx 0$, but nuclear knockout measurements completed the plot for strongly negative and positive values. It was found that when removing loosely bound particles the reduction factor was closer to one, while when removing deeply bound particles the reduction factor could be as small as 0.25.

What is learnt from the analysis of the knockout data is that, for removing valence nucleons from closed shell nuclei, there is a considerable reduction
of the spectroscopic strength as compared to the independent particle model, and that most large scale shell
model calculations account for only part of this reduction. This is a clear demonstration
of the key role of NN correlations within the nucleus. The exact nature of these
correlations is still not fully understood as well as its dependence on proton/neutron asymmetry
(see \cite{barbieri09,timofeyuk09} for two recent contributions on this topic).

Motivated by the exciting results from nuclear knockout experiments,
transfer reactions $^{34,36,46}$Ar(p,d)  were performed at  33 MeV/u,  with the intention of studying
the dependence of the reduction factor on the asymmetry $R(\Delta S)$. Results published in \cite{ar-exp}
show a clear disagreement on the trend of the reduction factor: while in knockout experiments one observes a strong
variation of the reduction factor, in transfer data no such dependence is present.
It has been pointed out that different beam energies could be at the heart of the disagreement since
nuclear knockout experiments were all performed at around 70 MeV/u
while the transfer measurement were done at 33 MeV/u. Consequently, the  $^{34,36,46}$Ar(p,d) experiments
have been repeated at higher energy and the analysis is in progress. Despite this aspect,
from the theoretical point of view,
it is pertinent to inspect the reaction theory used in the analyzes and evaluate the uncertainties.

Due to the renewed experimental interest, there have been a number of studies on reaction theory
for (d,p) and (p,d), particularly concerned with uncertainties in the methods. Examples include
the study of optical potential uncertainties \cite{liu04},
the effect of the reaction mechanism \cite{delaunay}, the uncertainties in the single particle
state \cite{muk05,pang06},  the description of the deuteron continuum \cite{moro09} and
the inclusion of full three-body dynamics in the reaction \cite{deltuva07}.

As in previous systematic studies \cite{lee07}, the recent $^{34,36,46}$Ar(p,d) data \cite{ar-exp} were analyzed using
the adiabatic wave model (ADWA) developed by Johnson and Soper \cite{soper}. This model provides an improvement
over distorted wave Born approximation (DWBA) in that it takes deuteron breakup into account explicitly,
however it also makes a zero-range approximation, reducing the accuracy of the method. In \cite{ar-exp}, a correction
to the zero-range approximation is introduced, namely the local energy approximation (LEA) \cite{lea}.
Note that, although not as simple as the zero-range ADWA, there is a finite range formulation of ADWA \cite{johnson-ria,tandy}
which has recently been studied in detail \cite{nguyen10} and compared to its zero range counterpart.
In \cite{nguyen10} finite range effects are shown to be important for (d,p) reaction at deuteron energies around
$50$ MeV or larger and the LEA prescription is shown to be inaccurate for energies well above the Coulomb barrier.
While the adiabatic model used in \cite{ar-exp} is very appealing due to its simplicity and the possibility of extracting
a single spectroscopic factor, one should remember that it presents an approximate formulation
to the full three-body problem $p+n+A$. The ADWA formulation is best if only small n-p distances contribute to the
transfer cross section. Faddeev techniques used to solve the exact three-body problem
are useful to estimate the uncertainty involved in the practical ADWA methods.

In this work we will re-analyze the results from \cite{ar-exp} with the finite-range version of
ADWA and using this improved reaction theory obtain the reduction factor for $^{34,36,46}$Ar (Section II).
We will also estimate uncertainties associated with the reaction theory, namely in what pertains
the interaction and in the three-body reaction dynamics (Section III). A final discussion and conclusions
will be presented in Section IV.

% ----------------------------------------------------------------------------------------
\section{Extracting spectroscopic factors from the Ar(p,d) data}
\label{sf}

\subsection{Theory}
\label{finite-range}

The most recent analysis on A(p,d)B reactions \cite{tsang05,lee06,lee07,tsang09,lee09}
have relied on the Johnson and Soper reaction model \cite{soper}. Given the relative importance of deuteron breakup
in (d,p) and (p,d) reactions, Johnson and Soper consider
the three-body problem of $n+p+B$ to describe the process.
The exact T-matrix for the process can be written in prior form as \cite{book}:
\begin{equation}
T = \langle \Psi^{(-)}| V_{np} + U_{pB} - U_{pA} |  \phi_{nB}\chi^{(+)}_{pA} \rangle \, ,
\label{texact-eq}
\end{equation}
where $\phi_{nB}$ is the initial bound state of the nucleus of interest $A$,
$\chi^{(+)}_{pA}$ describes the relative motion of the proton and the target
in the initial channel distorted by an optical potential $U_{pA}$, and $\Psi^{(-)}$
is the exact three-body wave function. Johnson and Soper expand the exact three-body
wave function in a complete set of eigenstates of the Hamiltonian of the $n+p$ subsystem:
\begin{equation}
\Psi^{(+)}(\vec r,\vec R)=\phi_d(\vec{r})\chi_d(\vec{R})+\int \mathrm{d}\vec{k}\phi^{(+)}_k(\vec{r}) \chi_k(\vec{R}) \;,
\label{phid-eq}
\end{equation}
which includes the ground state of the deuteron  $\phi_d(\vec{r})$ and all scattering states $\phi_k(\vec{r})$ of the $n+p$ system.
The Jacobi coordinates $(\vec r, \vec R)$ are the vector connecting the neutron and the proton, and the center of
mass of nucleus B and the center of mass of the deuteron, respectively.
In Ref.\cite{soper}, the zero-range approximation is made $V_{np}(r) \phi_d(r) = D_0 \delta(r)$.
With this approximation, the three-body problem for calculating $\Psi^{(+)}(\vec r,\vec R)$
reduces to the solution of an optical-model-like equation
where the distorting potential is the sum of the nucleon-target potentials
evaluated at half the deuteron energy $U_{ZR}(R)=U_{nA}(R)+U_{pA}(R)$:
\begin{equation}
(T_R + U_{ZR}(R) - E -  \epsilon_d ) \chi^{ZR}_d(\vec{R})=0 \, .
\label{js-eq}
\end{equation}
Here $T_R$ is the kinetic energy operator associated with $\vec R$.
This distorting potential $U_{ZR}$ incorporates deuteron breakup effects within the
range of $V_{np}$ and thus is not meant to describe deuteron elastic scattering.

In the Johnson and Soper model, the remnant term $(U_{pB} - U_{pA})$ in Eq.\ref{texact-eq} is neglected
and the transition amplitude simplifies to:
\begin{equation}
T = D_0 \langle \chi_{d}^{ZR(-)}(\vec R)|  \phi_{nB}(R) \chi^{(+)}_{pA}(R') \rangle \, ,
\label{texact-zr}
\end{equation}
where $\vec R' = \frac{m_B}{(m_B+m_n)} \vec R$.

It is well understood that generally the zero-range approximation for the deuteron
is not accurate. A full finite range version of the adiabatic model of \cite{soper} was introduced
by Johnson and Tandy \cite{tandy}. Therein the full three-body wave function is expanded in Weinberg states of the
$n+p$ system, and the Schr\"odinger equation for the three-body problem is reduced to a set of coupled channel equations.
The full coupled channel equations can be solved exactly  \cite{laid} but simplify tremendously when only the first
term of the Weinberg expansion is required.  Nguyen et al. \cite{nguyen10} use this approach
to perform a systematic study of the effects of finite range, involving a wide range of targets and beam energies.
Those results show that finite range effects can be very significant at intermediate energies.
A popular finite-range correction, the so-called local energy approximation (LEA), was introduced by Buttle and Goldfarb \cite{lea}
and relies on the truncation to first order of a series dependent on incoming momentum. Comparisons performed in \cite{nguyen10}
show that the local energy approximation performs well for sub-Coulomb energies but may fail at intermediate energies.

Here we follow the procedure in \cite{nguyen10} and evaluate the deuteron distorting potential,
taking into account the finite range of the deuteron:
\begin{equation}
U_{FR}(R)=  \frac{\langle \phi_d(\vec{r}) | V_{np} (U_{nA}+U_{pA}) | \phi_d(\vec{r}) \rangle}{\langle \phi_d(\vec{r}) | V_{np} | \phi_d(\vec{r}) \rangle} \, .
\label{jtpot2-eq}
\end{equation}
Once the adiabatic deuteron wave ${\tilde \chi}_{dB}$ is calculated using $U_{FR}(R)$, the  transfer amplitude can be obtained from:
\begin{equation}
T = \langle \phi_d {\tilde \chi}_{dB} | V_{np} |  \phi_{nB}\chi^{(+)}_{pA} \rangle \, .
\label{texact2-eq}
\end{equation}
where the remnant term has been neglected $(U_{pB} - U_{pA})$. Remnant term contributions for the cases under study here are no larger than $2\%$.

% --------------------------- RESULTS -----------------------------------

\subsection{Results}
\label{results}

We perform finite range calculations for the three reactions under study:
$^{34,36,46}$Ar(p,d)$^{33,34,45}$Ar(g.s.) at 33 MeV/nucleon as in \cite{nguyen10}.
For comparison with \cite{ar-exp}, we include results for the zero-range calculations
and the local energy approximation.
For the description of the Ar bound states of interest, we use a Woods Saxon mean field
with radius $r=1.25$ fm and diffuseness $a=0.65$ fm as in \cite{ar-exp}, plus a spin-orbit
interaction $V_{so}=6$ MeV with the same geometry as the mean field (note that the spin-orbit force
in the bound state does not affect the extracted spectroscopic factors).
Details on the single-hole bound states can be found in Table \ref{single}.
\begin{table}[htdp]
\caption{Properties of the single hole states: $S_n$ is the separation energy,
$R_{rms}$ is the r.m.s. radius of the hole state, and $b_{nlj}$ is it asymptotic
normalization constant \cite{muk05}.}
\label{single}
\begin{center}
\begin{tabular}{|c|c|c c c|}
\hline
nucleus         & state &  $S_n$ (MeV) & $R_{rms}$ & $|b_{nlj}|$ \\
\hline
$^{33}$Ar  	& $2s_{1/2}$&	17.07  & 3.39 	& 31.8 \\
$^{35}$Ar  	& $1d_{3/2}$&	15.25  & 3.40 	& 10.3 \\
$^{45}$Ar  	& $1f_{7/2}$&	 8.07  & 4.10 	& 2.35 \\
\hline
\end{tabular}
\end{center}
\end{table}

\begin{figure}[t!]
{\centering \resizebox*{0.425\textwidth}{!}{\includegraphics{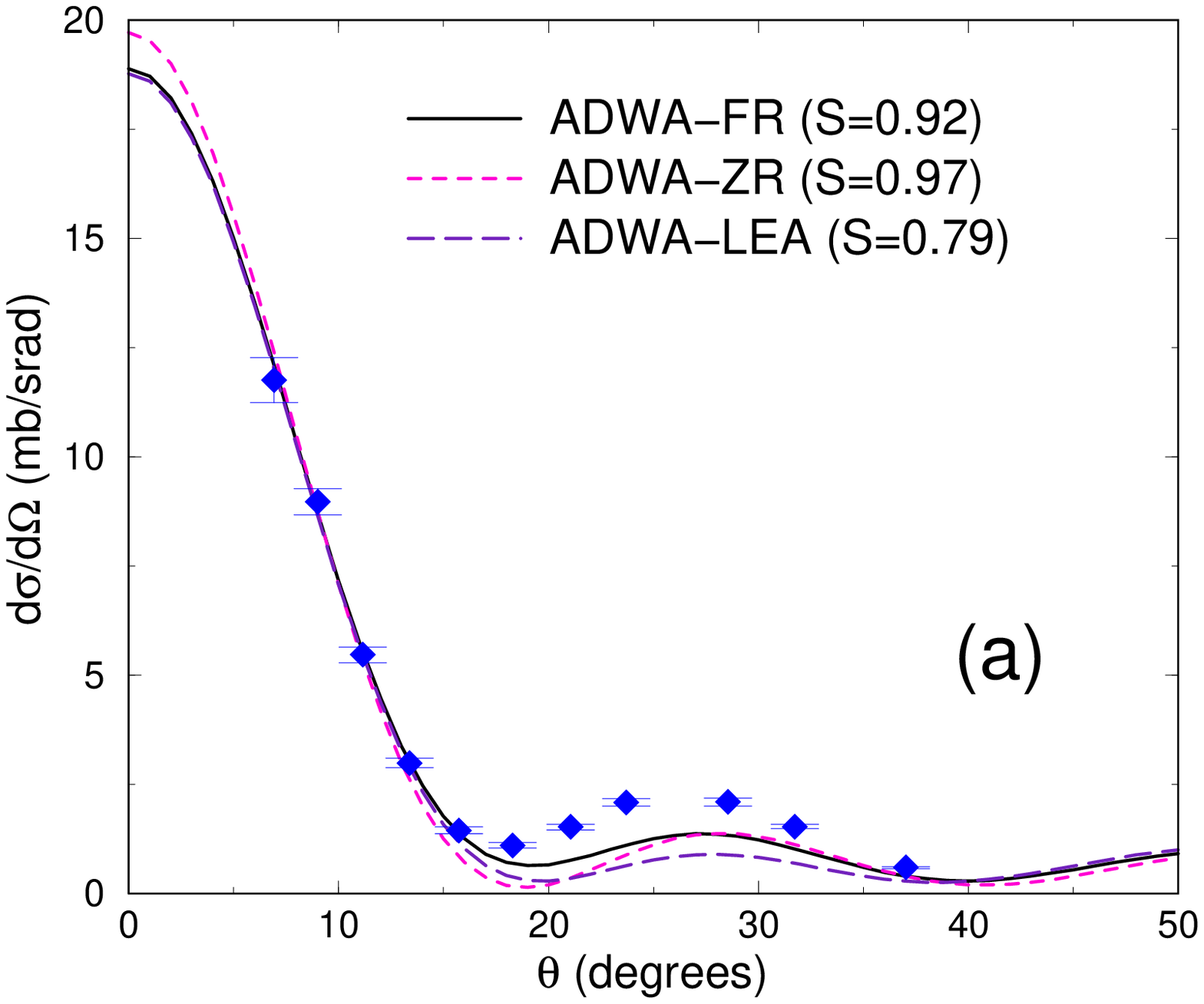}}} \\
{\centering \resizebox*{0.425\textwidth}{!}{\includegraphics{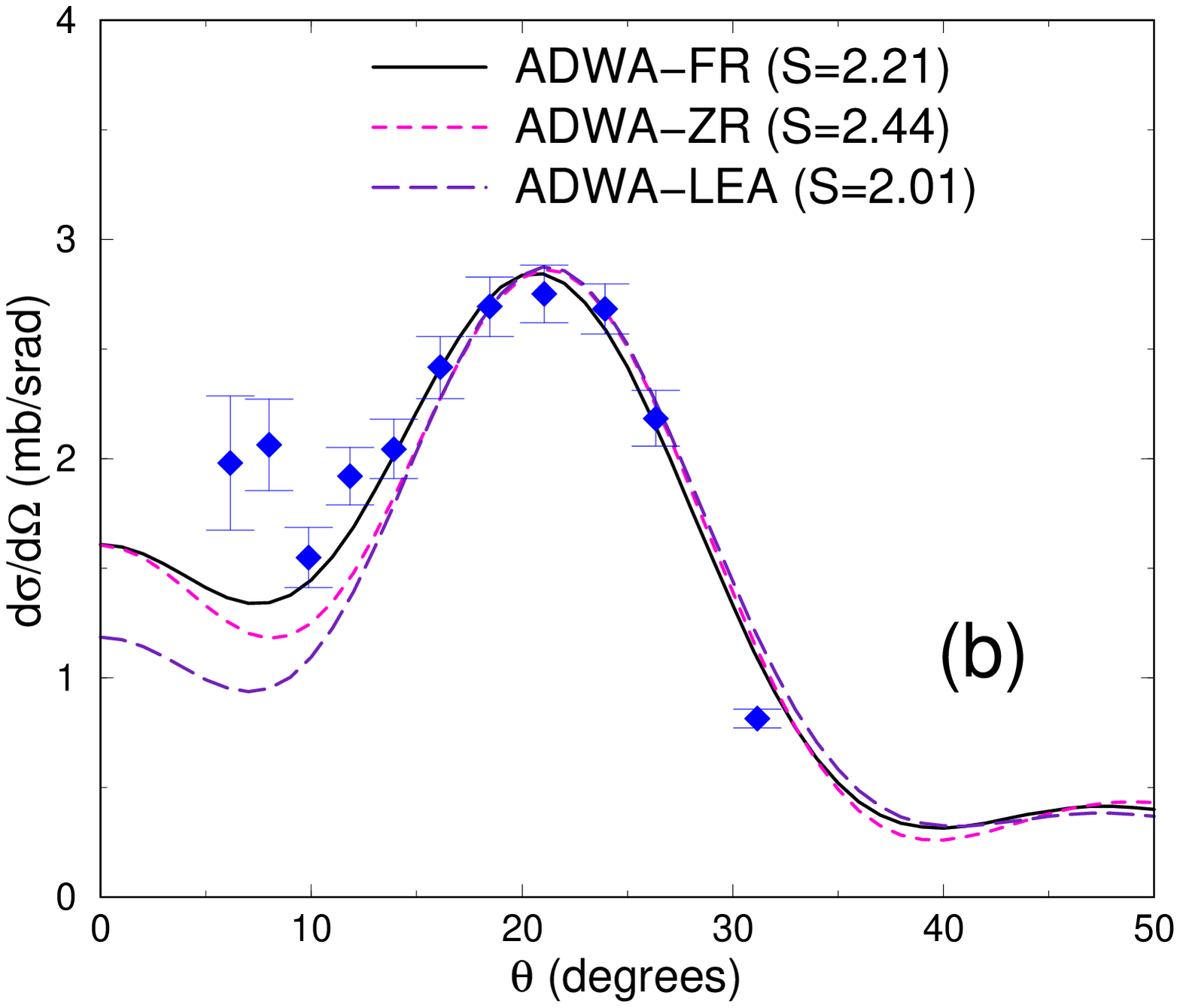}}} \\
{\centering \resizebox*{0.425\textwidth}{!}{\includegraphics{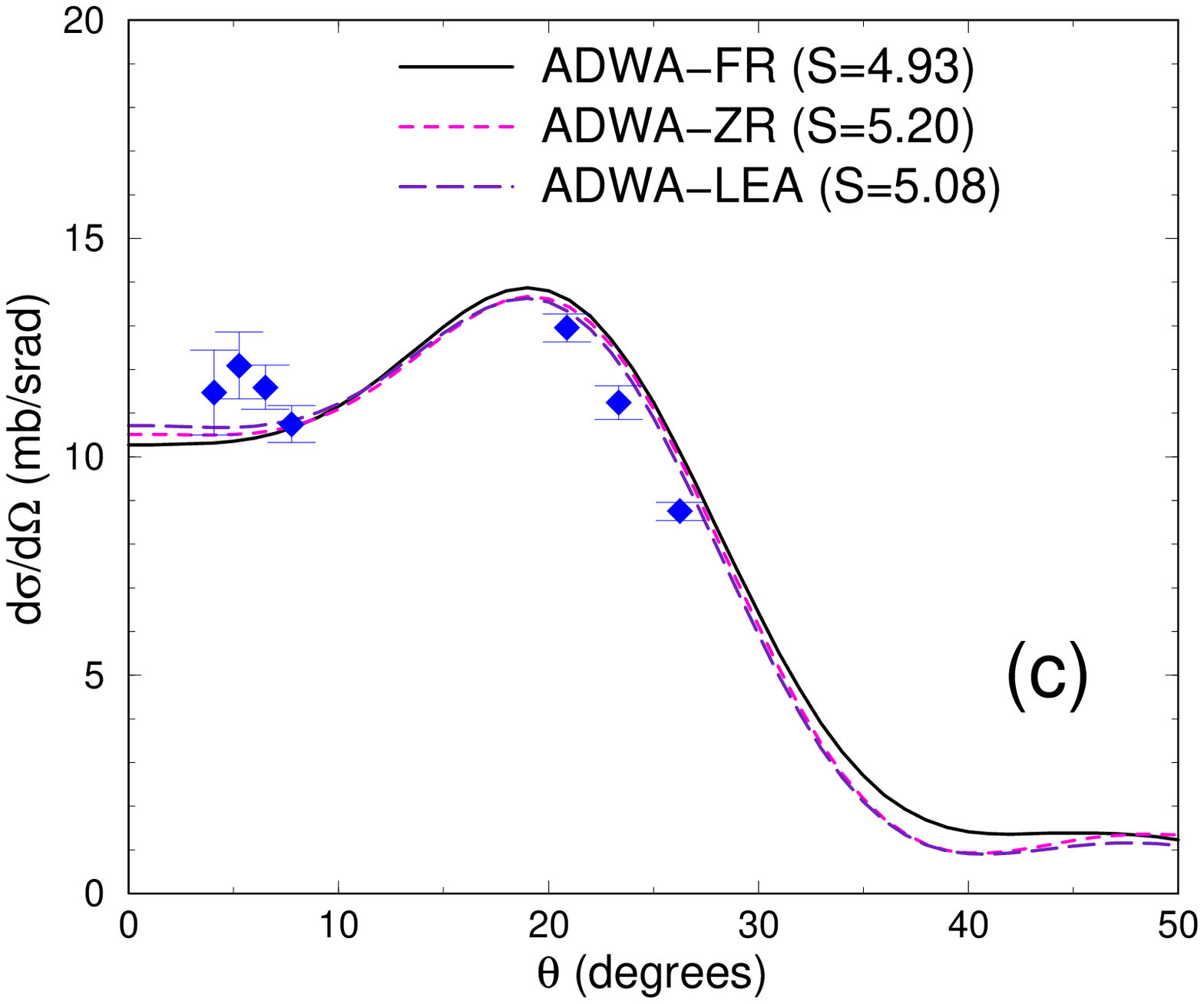}}}
\caption{\label{xs-lin} (Color online) Angular distributions for:
(a) $^{34}$Ar(p,d)$^{33}$Ar(g.s.) $E_p = 33$ MeV,
(b) $^{36}$Ar(p,d)$^{35}$Ar(g.s.) $E_p = 33$ MeV and
(c)  $^{46}$Ar(p,d)$^{45}$Ar(g.s.) $E_p = 33$ MeV.
Comparison of full finite range (solid) with the zero-range approximation (dashed),
and the local energy approximation (long-dashed). All distributions have
been multiplied to scale the data by the indicated spectroscopic factor $S$. }
\end{figure}

For the nucleon optical potentials we use Chapel Hill \cite{ch89} as in \cite{ar-exp}
unless otherwise stated. Finally, for the deuteron bound state and the $V_{np}$ interaction
appearing in the transfer operator of Eq.(\ref{texact2-eq}), we use the Reid interaction
\cite{rsc} and the corresponding $D_0$ in the zero range model. As stated earlier, the
remnant term is neglected but was determined to be less that $2\%$ in the cases
here studied. The finite range deuteron adiabatic potentials of Eq.(\ref{jtpot2-eq}) are generated using the code
{\sc twofnr} \cite{twofnr} and the transfer cross sections are computed with the code {\sc fresco} \cite{fresco}.

Full finite range calculations (solid line) are compared to the zero-range approximation (dashed line) and the local-energy approximation (long-dashed line), together with the measured angular distributions in Fig.\ref{xs-lin}. All calculations have been renormalized to the data using
the same procedure as in \cite{ar-exp}: the first 3 data points are used for $^{34}$Ar, the 5 points around
the peak are used for $^{36}$Ar and all points but the first are used for $^{46}$Ar.
The extracted spectroscopic factors $S$ are provided in the legend of Fig.\ref{xs-lin}. Overall the angular distributions are well described
by theory. It is only for the $^{36}$Ar case that we find finite range effects to introduce differences
in the shape of the distribution. For this case these effects
improve the description of the data. Most importantly, the normalization of the
cross section changes significantly with finite range effects.
As compared to cross sections calculated including full finite range effects, zero range results
systematically overestimate the cross section: $5$\% for $^{34}$Ar(p,d)$^{33}$Ar, $9$\% for $^{36}$Ar(p,d)$^{35}$Ar
and $5$\% for $^{46}$Ar(p,d)$^{45}$Ar.
On the other hand, LEA results
underestimate the cross section by $14$\% for $^{34}$Ar(p,d)$^{33}$Ar and $9$\% for $^{36}$Ar(p,d)$^{35}$Ar
while overestimating the cross section by $3$\% for $^{46}$Ar(p,d)$^{45}$Ar.

\begin{table}[t!]
\caption{Extracted spectroscopic factors using the finite-range ADWA model SF(ADWA-RF),
and the percentage differences with results obtained with the zero-range
approximation, the local energy approximation, and the inclusion of the remnant term.
Also shown are the spectroscopic factors calculated in large scale shell model SF(LSSM) \cite{ar-exp}.
}
\label{effects}
\begin{center}
\begin{tabular}{|c| c c c |}
\hline
Model  & $^{34}$Ar &  $^{36}$Ar & $^{46}$Ar \\
\hline
SF(ADWA-FR)& 0.92	&	2.21	&	4.93 \\
SF(LSSM) &	 1.31 	&	2.10	&	5.16 \\
\hline
zero range  	   &	5\%	&	9\%	&	5\% \\
local energy approx.  &	14\%	&	9\%	&	3\% \\
remnant &	          2\%	&	0.3\%	&	0.6\% \\
\hline
\end{tabular}
\end{center}
\end{table}

\section{Estimating errors from reaction theory}

The finite range adiabatic model is by no means the full solution and an important
part of the analysis is to understand the uncertainties introduced with the
several approximations made. The most important approximation of course is
that we transform the many-body problem into a three-body problem, which introduces effective
interactions and with them uncertainties. However, there are also approximations at the three-body level,
namely the truncation of the expansion in Weinberg states as well as not taking into account
the coupling between breakup and transfer channels to all orders. The exact solution to the full three-body
can be obtained in the Faddeev framework. Below we present some of the essential features of the
Faddeev method in momentum space and follow with a discussion on the estimation of reaction
theory uncertainties for the problem.

\subsection{Faddeev theory}
\label{faddeev}

Recently, nuclear reactions involving protons or deuterons have been studied solving the full
Faddeev-type equations in momentum space \cite{deltuva09}.
These equations, usually referred to as AGS equations (for Alt, Grassberger and Sandhas) \cite{Alt67},
are coupled integral equations for the transition operators
\begin{equation}\label{eq:AGS}
 U_{\beta \alpha} = {} (1 - {\delta}_{\beta \alpha}) G_0^{-1} %\nonumber \\
     + \sum_{\sigma=1}^{3} (1 - {\delta}_{\beta \sigma}) T_\sigma  G_0 U_{\sigma \alpha}
\end{equation}
where $\alpha,\beta$ refer to the three Faddeev components, which also denotes
the associated two-body pairs. The on-shell matrix elements
$\langle\psi_{\beta}|U_{\beta \alpha}|\psi_{\alpha}\rangle$
are scattering amplitudes and therefore lead directly to the  observables.
In Eq.~(\ref{eq:AGS}) $G_0 = (E+i0-H_0)^{-1}$ is the free resolvent, $E$ being the available
three-particle energy and $H_0$ the free Hamiltonian.
The  two-particle transition operators $T_{\sigma}$ are obtained from the Lippmann-Schwinger
equation
\begin{equation}
T_{\sigma} = v_{\sigma} + v_{\sigma} G_0 T_{\sigma}
\end{equation}
where $v_{\sigma}$ is the potential for the pair $\sigma$ in the odd-man-out notation.
Thus, AGS equations may be viewed as a way for summing up the multiple scattering series in terms
of  $T_{\sigma}$.
Each Faddeev channel state $|\psi_{\beta}\rangle$ for  $\beta = 1,2,3$ is an
eigenstate of the channel Hamiltonian $H_\beta = H_0 + v_\beta$
with the energy eigenvalue $E$; thus, $|\psi_{\beta}\rangle$ is a product of
the bound state wave function for pair $\beta$ and a plane wave
with fixed on-shell momentum
corresponding to the relative motion of particle $\beta$ and pair $\beta$
in the initial or final state. The channel states $|\psi_{0}\rangle$
are the eigenstates of $H_0$ with the same eigenvalue $E$ and
describe the free motion of three particles.
Observables of elastic scattering are calculated from the matrix elements with
$\beta = \alpha$, those of breakup are given by $\beta=0$
 while $0 \neq \beta \neq \alpha$ correspond to transfer reactions.
Although the original AGS equations were derived for short-range potentials, the Coulomb interaction can be included
using the method of  screening and renormalization \cite{deltuva05}.
The numerical details of solving AGS equations with potentials of general form
can be found in \cite{deltuva:phd,deltuva:03a}.
Since elastic scattering, transfer, and breakup
are treated on equal footing, the Faddeev/AGS framework yields the most accurate solution to the general three-body
problem where all channels (elastic, transfer and breakup) are fully coupled; it already
has provided important tests to other methods commonly used in reaction theory \cite{deltuva07,crespo07,crespo08}.

%{\bf Arnas, i would like to have a more intuitive description
%of the AGS equations. could you provide a couple of sentences explaining the relations above?}.

\subsection{Discussion on uncertainties from reaction theory}
\label{discussion}

Although the Faddeev/AGS framework offers an exact solution for elastic scattering, transfer and breakup,
given a fixed three-body Hamiltonian,
it does have several conceptual limitations regarding applications to nuclear reactions because these are not
genuine three-body problems.
A first one is related to the mapping onto the many-body problem.
As pointed out in \cite{timofeyuk99}, one of the important features of the transition amplitude for the
transfer process where only $V_{np}$ appears in the transfer operator is that the mapping onto the many-body problem can
be done in a rather  straight-forward manner in terms of one-nucleon overlap functions. In the Faddeev approach
this is no longer possible because elastic, breakup and transfer components are all mixed. This implies
that the Faddeev approach to (p,d) reactions should only be used for pure single particle states of nucleus
$A$ and cannot be used to extract a spectroscopic factor from a ratio to experimental cross sections \footnote{
Alternative ways to determine spectroscopic factors from the many-body problem are being explored \cite{timofeyuk09}}.
In this work we will use the Faddeev solutions to estimate the theoretical errors in the finite range
adiabatic model only.

Second, the rigorous Faddeev theory can only use the same potentials in the initial and final states,
while in the adiabatic approaches for A(p,d)B reactions \cite{soper,tandy,nguyen10} the
nucleon-nucleus interaction $V_{NB}$ is different in the initial and final channel.
% based on the understanding of the physics of the problem in the Faddeev framework the nucleon-nucleus potentials are fixed.
Within the adiabatic approaches, for the neutron in the initial channel we use
an effective real binding potential fitted to the spectrum of nucleus $A$ whereas in the final state
the nucleon optical potential contains absorption which reproduces elastic scattering.
In previous Faddeev A(p,d)B calculations $V_{nB}$ was real in all partial waves \cite{deltuva07,deltuva09}
while $V_{pB}$ was complex with parameters corresponding to the $A+p$ state.
In the present work, Faddeev calculations take $V_{nB}$ to be real in the partial wave with the
$n+B$ bound state but in all other partial waves $V_{nB}$ is taken to be complex
as in the adiabatic approach in the final state.
The effect of neutron absorption in the partial wave with the $n+B$ bound state
can be estimated by introducing energy  dependent interactions into the AGS equations \cite{deltuva-E}.
Even if not fully consistent, we include these calculations here to provide an estimate of the magnitude of the effect
of energy dependence in the neutron-Ar interaction.
Within adiabatic approaches, the proton interaction in the initial channel $U_{pA}$ is calculated
at the beam energy, whereas in the exit channel $U_{pB}$ it is calculated at half the deuteron energy.
In Faddeev calculations, the proton interaction $U_{pB}$ is fixed at the beam energy.

There are also some technical challenges in the Faddeev/AGS framework  regarding applications to nuclear reactions,
related to the convergence of multiple scattering series,  partial wave expansion and Coulomb screening.
However, these difficulties do not preclude us from obtaining well converged results for the reactions
considered in the present work.

Finally, another important consideration in determining the uncertainties in reaction theory relates
to the optical potentials. The finite range adiabatic theory is built on nucleon optical potentials which
are much better known than deuteron optical potentials. Nevertheless these are not uniquely determined
and thus, in addition to the reference Chapel Hill potential \cite{ch89}, we also performed calculations
with a more recent global potential by Koning and Delaroche \cite{koning}. From the percentage
difference in cross sections obtained with these two global potentials, we estimate the
optical potential uncertainty in the theoretical cross section.

\subsection{Estimates of the uncertainties from reaction theory}

\begin{figure}[t!]
{\centering \resizebox*{0.425\textwidth}{!}{\includegraphics{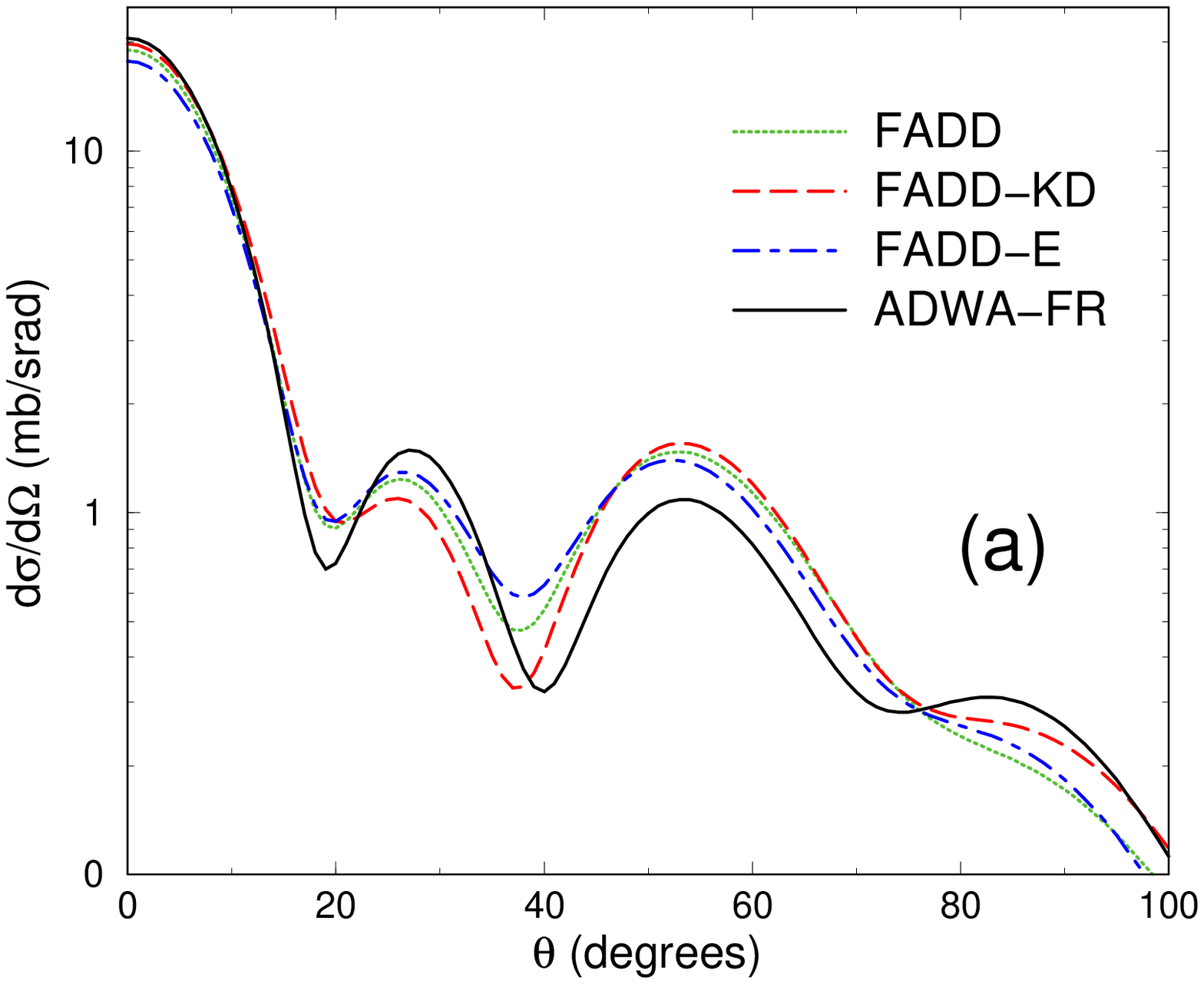}}} \\
{\centering \resizebox*{0.425\textwidth}{!}{\includegraphics{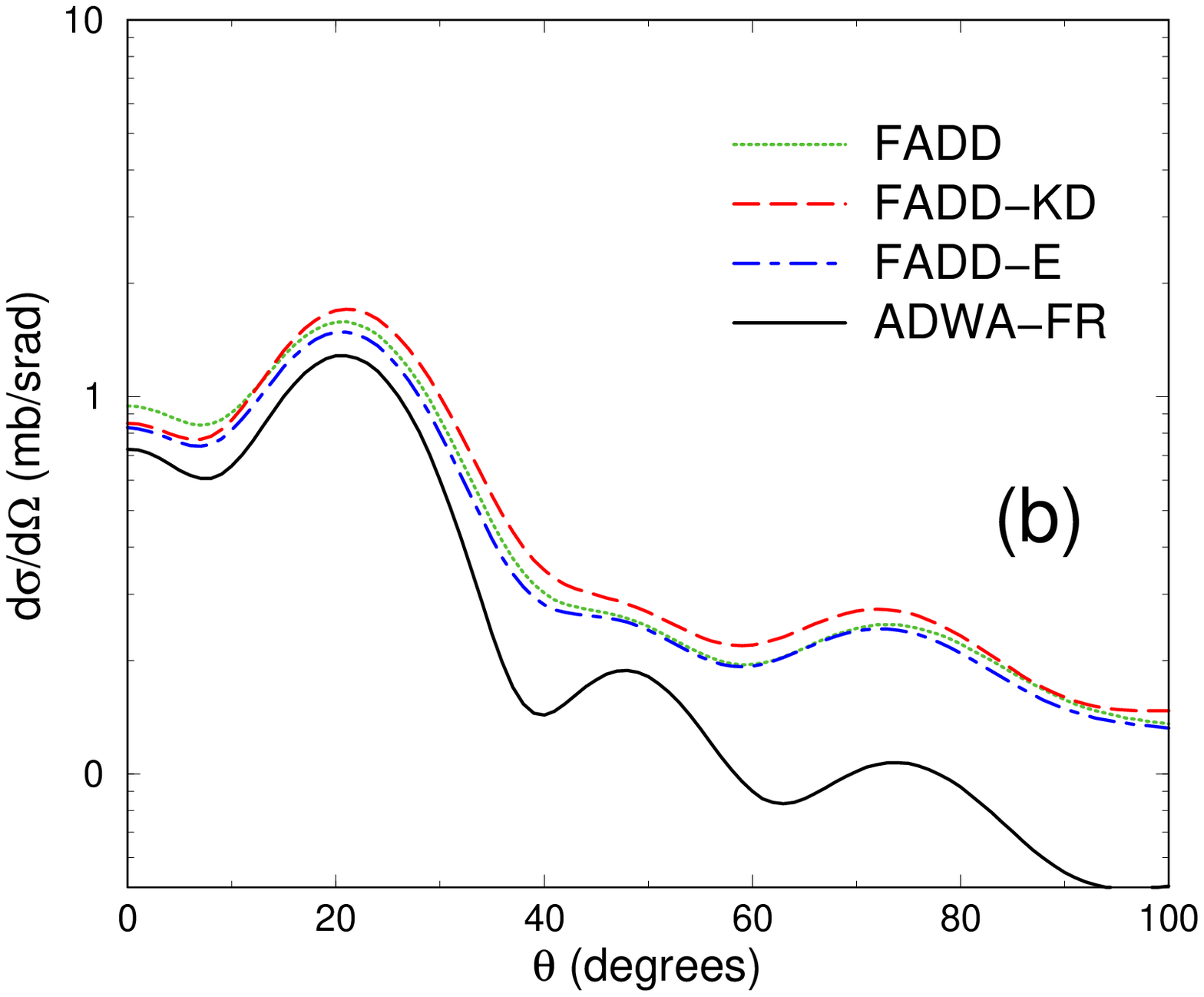}}} \\
{\centering \resizebox*{0.425\textwidth}{!}{\includegraphics{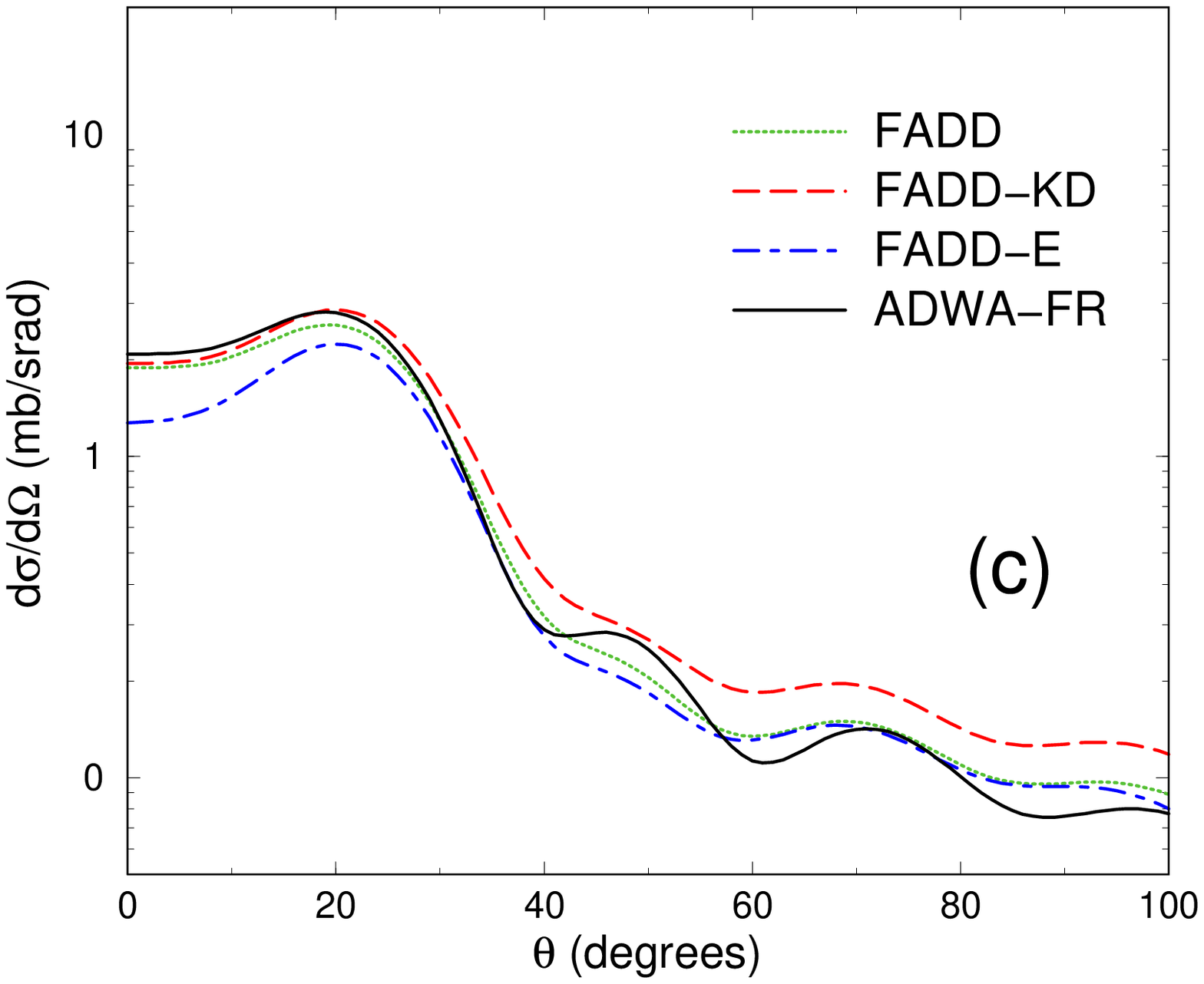}}}
\caption{\label{xs-log} (Color online) Angular distributions for:
(a) $^{34}$Ar(p,d)$^{33}$Ar(g.s.) $E_p = 33$ MeV,
(b) $^{36}$Ar(p,d)$^{35}$Ar(g.s.) $E_p = 33$ MeV and
(c)  $^{46}$Ar(p,d)$^{45}$Ar(g.s.) $E_p = 33$ MeV.
Full three-body Faddeev calculations (FADD) with the finite range adiabatic model (ADWA-FR).
More detail in the text. }
\end{figure}
In order to estimate the error from reaction theory, full three-body Faddeev calculations were performed
using the same interactions, as discussed in Section \ref{results} and \ref{discussion}. The results from these calculations (FADD)
are compared to the finite range adiabatic calculations (ADWA-FR) in Fig. \ref{xs-log}. The solid green lines correspond
to the Faddeev results, while the solid black lines correspond to the finite range adiabatic.
In these plots, no normalization to the data is performed. Percentage differences between cross sections
obtained within the finite range adiabatic  model and the Faddeev equations are calculated relative to the
Faddeev results. These comparisons are performed in the angular region
where spectroscopic factors were extracted from the data \cite{ar-exp} and are used to estimate the error of
the approximations in the reaction mechanism. The relevant angles are: $\theta\approx 9 ^\circ$ for $^{34}$Ar,
$\theta \approx 20 ^\circ$ for $^{36}$Ar and $\theta \approx 5 ^\circ$ for $^{46}$Ar.
The percentage differences are presented in Table \ref{errors}. In addition, we also perform
Faddeev calculations where instead of using the Chapel Hill nucleon potential, we use the more modern Koning
and de la Roche \cite{koning}. The resulting angular distributions (FADD-KD) correspond to the dashed red lines in
Figs. \ref{xs-log}. Percentage differences between these two Faddeev results in the same angular range
of interest are presented in Table \ref{errors}. For completeness, and to ensure that our error estimates
in Table \ref{errors} are reliable, we also perform Faddeev calculations including the imaginary term
of $V_{nB}$ also in the partial wave with the  $n+B$ bound state, for all $E_{nB}>0$ \cite{deltuva-E}.
Results (FADD-E) are shown in Fig. \ref{xs-log}
as the blue dashed lines and we find that differences between these two Faddeev results are of the same
magnitude as the errors estimated when the $V_{nB}$ is real only. Since there are ambiguities when introducing
the energy dependence of $V_{nB}$, we do not use FADD-E to estimate errors.

\begin{table}[t!]
\caption{Estimates of theoretical errors in the extracted spectroscopic factors
due to approximations in the reaction model as well as experimental errors.}
\label{errors}
\begin{center}
\begin{tabular}{|c|ccc|}
\hline
Errors  & $\epsilon_{th}$($^{34}$Ar) &  $\epsilon_{th}$($^{36}$Ar) & $\epsilon_{th}$($^{46}$Ar) \\
\hline
Optical potential  &	8 \%	&	7\%	    &	4\% \\
Faddeev  &	            6 \%	&	19\%	&	11\% \\
Experiment &            8\%    &     8\%    &   8\%  \\
\hline
Total &		13 \%	&	22 \%	&	14 \% \\
\hline
\end{tabular}
\end{center}
\end{table}

As the optical potential uncertainty is independent of the three-body effects included in the
Faddeev description, and these are both independent from the experimental statistical error,
we estimate the total error by summing in quadrature. These are shown in the last
line of Table \ref{errors}. The largest increase in error is seen for the $^{36}$Ar case,
and originates from the approximations in the reaction mechanism.

\begin{figure}[t!]
{\centering \resizebox*{0.42\textwidth}{!}{\includegraphics{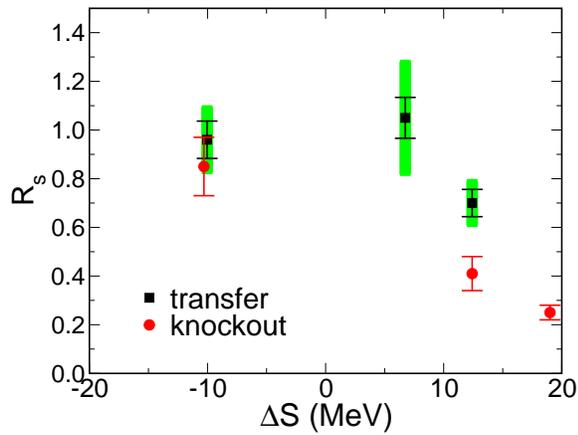}}}
\caption{\label{reduction} (Color online) Reduction factors $Rs=SF(ADWA-FR)/SF(LB-SM)$ as a function of the difference between
the neutron and proton separation energies $\Delta S$. The  squares and  circles correspond to values
extracted using transfer or knockout respectively. The bars correspond to the total uncertainty including
both experimental and theoretical errors evaluated for the transfer reactions.}
\end{figure}

Finally, as in previous studies \cite{ar-exp}, we calculate the reduction factor $R_s$ as a ratio of the experimentally
determined spectroscopic factor and that obtained from large basis shell model calculations.
For the purpose of consistency, we use the same shell model calculations reported in \cite{ar-exp}.
These include the sd-shell model space and the USDB effective interaction \cite{Sig07} for $^{34,36}$Ar
and the sd-pf model space with the interaction of Nummela  et al. \cite{Bro06} for $^{46}$Ar.
The predicted ground state spectroscopic factors for $^{34}$Ar, $^{36}$Ar and $^{46}$Ar are
$1.31$, $2.10$ and $5.16$, respectively, as shown in Table II. The reduction factors for $^{34}$Ar, $^{36}$Ar and $^{46}$Ar
using the spectroscopic factors extracted from transfer cross sections and finite range adiabatic theory
are plotted as black squares in Fig. \ref{reduction} and errors bars include only the experimental error.
The total errors, including those from Table \ref{errors}, are represented by the green bars.
The knockout results \cite{ar-exp} are plotted as red circles and only include statistical errors.

\section{Conclusions}

In summary, we have reanalyzed the reactions  $^{34,36,46}$Ar(p,d)$^{33,35,45}$Ar at 33 MeV/nucleon
using an improved reaction theory which includes deuteron breakup and finite range effects,
based on the adiabatic wave approximation (ADWA).
In addition we have quantified the errors in the reaction theory due to the
optical potential and the approximate solution of the three-body problem. In order
to do this we have performed exact three-body Faddeev calculations.
Since our theory is based on nucleon optical potentials,
we find that the optical potential uncertainty is below $10$\%.
However, differences between the adiabatic results and the Faddeev results can be as
large as $\approx 20$\%.
Further detailed studies to better understand the source of these differences are underway.

With the present level of accuracy of the reaction theory, the slope suggested by the knockout
data can be corroborated by the transfer data.
Our error bars do not include the uncertainties in the single neutron-hole overlap functions.
Here we use standard geometry and it is not clear whether, specially for $^46$Ar, this will be
an adequate choice. Measurements of the matter radius could provide additional constrains.
Ideally, these reactions could be repeated at lower energy in order to extract
the asymptotic normalization of the overlap functions, which would then enable a
much better handle on this additional ambiguity \cite{muk05}.

Although in this work we concentrate on the reaction theory for transfer, it is equally important to
estimate theoretical errors associated with the description of the knockout reactions.
Our results call for a better understanding of the reaction mechanism in order to reduce the errors.

We are grateful to Jenny Lee and Betty Tsang for providing the data and many useful discussions.
This work was partially supported by the National Science Foundation
grant PHY-0555893, the Department of Energy through grant DE-FG52-08NA28552
and the TORUS collaboration DE-SC0004087 and the Portuguese Foundation of Science
and Technology PTDC/FIS/65736/2006.

%\end{document}

\end{document}